\documentclass[a4paper,11pt]{article}
\usepackage{pos}
\usepackage{float}
\usepackage{enumitem}

\definecolor{hookgreen}{rgb}{0.0,0.44,0.0}

\newlength{\bibitemsep}
\newlength{\bibparskip}
\let\oldthebibliography\thebibliography
\renewcommand\thebibliography[1]{%
  \oldthebibliography{#1}%
  \setlength{\parskip}{\bibitemsep}%
  \setlength{\itemsep}{\bibparskip}%
}

\title{Using the Cherenkov Telescope onboard EUSO-SPB2 for Target of Opportunity searches of very high energy neutrino sources}
\ShortTitle{Using the Cherenkov Telescope onboard EUSO-SPB2 for Target of Opportunity searches}

\author[a]{Tobias Heibges}
\author*[c,d]{Claire Gu\'epin}
\author[b]{Diksha Garg}
\author[b]{Luke Kupari}
\author[b]{Mary Hall Reno}
\author[e]{Tonia M. Venters}
\author[a]{Lawrence Wiencke}

\affiliation[a]{Colorado School of Mines, Department of Physics,\\
Golden, CO, USA}

\affiliation[b]{University of Iowa, Department of Physics and Astronomy,\\
Iowa City, IA,  USA}

\affiliation[c]{University of Chicago, KICP,\\
Chicago, IL, USA}

\affiliation[d]{Laboratoire Univers et Particules de Montpellier,\\
Montpellier, France}

\affiliation[e]{Goddard Space Flight Center,\\
Greenbelt, MD, USA}

\onbehalf{for the JEM-EUSO Collaboration}

\emailAdd{theibges@mines.edu}

\abstract{The Extreme Universe Space Observatory on a Super Pressure Balloon 2 (EUSO-SPB2) mission launched from Wanaka New Zealand on May 13, 2023. The mission ended after $36$\,h due to a balloon leak that resulted in the payload being lost in the Pacific Ocean. Over the course of the mission, the onboard Cherenkov Telescope (CT) was pointed just below the Earth's limb to search for optical signals from upward-moving extensive air showers that were induced by decaying tau-leptons that were generated by the conversion of $>10$ PeV tau-neutrinos in Earth. Such very-high energy (VHE) neutrinos may be produced in several classes of astrophysical sources that are suspected of possibly accelerating particles to ultra-high energies. In this contribution, we discuss the EUSO-SPB2 Target-of-Opportunity (ToO) campaign to search for VHE neutrino signals coming from astrophysical sources that crossed the CT's field of view ($6.4^\circ \times 12.8^\circ$). We also present a new software tool designed for scheduling observations, the Neutrino Target Scheduler, that we developed to support the ToO campaign. We also calculate upper limits on the neutrino fluences from sources observed during the ToO campaign. Our observations demonstrate the viability of conducting ToO follow-up observations from a near space environment with future balloon missions, such as the pathfinder mission POEMMA-Balloon with Radio.}

\ConferenceLogo{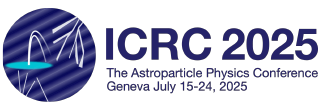}

\FullConference{%
39th International Cosmic Ray Conference (ICRC2025)\\
  15 - 24 July 2025\\
  Geneva, Switzerland}

\begin{document}
\maketitle

\section{Introduction}
\label{sec:intro}

Detecting high-energy (HE) neutrinos from astrophysical sources opens new avenues to understanding some of the most energetic processes in the universe, particularly cosmic particle acceleration up to ultra-high energies (UHE). Already, blazar flares and tidal disruption events  have what may be coincident HE neutrino events detected by IceCube \cite{IceCube:2018dnn, IceCube:2018cha, 2019ATel12967....1T, Giommi:2020viy,Stein:2020xhk,Reusch:2021ztx, vanVelzen:2021zsm}. The active galactic nucleus NGC~1068 has an associated HE neutrino flux \cite{IceCube:2022der}. The IceCube Collaboration has also reported the first observation of HE neutrinos associated with the Galactic Plane \cite{IceCube:2023ame}. Recently, KM3NeT reported the first very-high energy (VHE)\footnote{Defined as $E \gtrsim 10$~PeV.} neutrino detected with an estimated energy of hundreds of PeV \cite{KM3NeT:2025npi}. Balloon-borne and space-based Cherenkov telescopes have the potential to expand the detection of neutrino sources to VHEs
\cite{Venters:2019xwi, ICRC2023Heibges}. The technique relies on detecting Earth-skimming tau neutrinos that convert to tau leptons, which emerge from the Earth and produce upward-going extensive air showers (EASs) that produce Cherenkov radiation. To be observable, neutrino sources must dip below the Earth's limb, as viewed from the telescope, so that the Earth acts as a neutrino converter. 

The Extreme Universe Space Observatory on a Super Pressure Balloon 2 (EUSO-SPB2) mission, launched from Wanaka, New Zealand on May 13, 2023 carried a Cherenkov telescope (CT) designed to detect optical Cherenkov signals of upward-going EASs that originate from $>10$ PeV tau neutrinos \cite{Bagheri:2024byu,Adams:2025owi}. While super-pressure balloons are designed to allow flight durations up to $\sim 100$ days, a leak in the balloon required the termination of EUSO-SPB2 after only two nights of observation. Though the mission ended prematurely, Target-of-Opportunity (ToO) follow-up observations of candidate neutrino sources and the subsequent data analysis demonstrate the technique of searching for VHE tau neutrinos through the Cherenkov channel from a near-space environment. 

\section{Cherenkov telescope on EUSO-SPB2}
\label{sec:CT}

The EUSO-SPB2 CT was designed to detect beamed Cherenkov light produced by EASs initiated by either VHE neutrinos converting below the Earth's limb or HE cosmic rays interacting in the atmosphere above it. The CT featured a 1~m diameter entrance pupil and a camera consisting of 512 silicon photomultipliers (SiPMs) that covered a field of view (FoV) of $6.4^\circ$ vertically and $12.8^\circ$ horizontally \cite{Bagheri:2024byu,Adams:2025owi}. A bifocal Schmidt optical system was employed to reject direct cosmic hits. Due to the limited azimuth coverage of the camera, the CT could not instantaneously monitor the entire region below the limb from where detectable signals from VHE neutrinos may originate. To enable follow-up observations of transient events, the telescope was mounted on a NASA rotator that provided $360^\circ$ rotation in azimuth and a pointing accuracy of $\pm 5^\circ$. In the vertical direction, a linear stage enabled the CT to tilt the optical axis from $+3.5^\circ$ to $-12.5^\circ$ in elevation, enabling observation both above and below the limb.

Only a limited number of sources could be observed each night, constrained by a finite number of CT re-pointings and the restricted total observation time. Additionally, observations required dark-sky conditions -- specifically, the Sun had to be $> 18^\circ$ below the Earth's limb and the Moon had to be $< 5$\% illuminated or below the horizon. Finally, the already limited observation time was further constrained by pointings above the limb to conduct HE cosmic-ray observations. In light of these limitations, the EUSO-SPB2 ToO observation program was designed around scheduling observations of the most promising transient sources for VHE neutrino searches. For this purpose, we developed the Neutrino Target Scheduler (NuTS) software, which is described in the next section.

\section{Neutrino Target Scheduler (NuTS) software}
\label{sec:NuTS}

We designed the NuTS software to optimize ToO searches during the flight of EUSO-SPB2 and to support future missions such as POEMMA-Balloon with Radio (PBR) \cite{ICRC2025Eser}. This software is part of a full simulation chain developed by the JEM-EUSO Collaboration \cite{ICRC2025HeibgesDiffuse}. NuTS includes three separate modules:
\begin{itemize}[itemsep=0pt]
    \item A listener module that collects and parses alerts from the General Coordinates Network (GCN), the Transient Name Server (TNS), and other alert systems, and produces a database of transient and steady sources of interest for VHE neutrino searches; 
    \item An observability module that uses the source database, the locations of the Sun and the Moon, the Moon illumination, and flight information to compile a list of observable sources;
    \item A scheduler module that employs a  prioritization scheme (selected from several available options) on the list of observable sources to select the ones that should be prioritized.
\end{itemize}
Our prioritization scheme for the sources is based on flux predictions from models in the literature and the relative occurrence rate in the local universe \citep[e.g.,][]{Venters:2019xwi, Guepin:2022qpl, 2023ICRCWistrand}. Using this as a guide, sources are classified in different priority tiers. From higher to lower priority: 
1.\,rare, noteworthy galactic transients (e.g., a galactic supernova),
2.\,extragalactic binary neutron star mergers and IceCube gold or bronze events \cite{blaufuss2019generation},
3.\,blazar flares,
4.\,tidal disruption events (TDEs),
5.\,gamma ray bursts (GRBs),
6.\,active galactic nuclei (AGNs) other than blazars (alerts from TNS),
7.\,extragalactic supernovae (SNe),
8.\,steady sources compiled from the literature (e.g., nearby starburst galaxies and AGNs, cosmic-ray hotspots, etc.).
NuTS will soon be released as an open-source software, and more details about its functionalities can be found in \cite{Heibges:2025llt}.

\section{Target of Opportunity Follow-up Searches}\label{sec:target-limit-analysis}

\subsection{Source selection}

During EUSO-SPB2's ToO follow-up campaign, we compiled a comprehensive catalog of transient and steady sources. This catalog includes a large number of alerts, notably $1224$ GCN alerts and $670$ TNS alerts issued between May 2022 and May 14, 2023. Additionally, $16$ alerts from the Astronomer's Telegram (ATels) were added by hand to the database. The catalog also features $120$ steady sources, primarily selected from the TeV Gamma-Ray Source Catalog (TeVCat).

The list of transient sources was further refined by applying a time window for ToO follow-up searches, which varies depending on the type of transient. This is based on VHE emission models, which indicate that different types of transients are likely to produce VHE neutrinos over distinct timescales. For example, the follow-up time window for GRBs is set to a maximum of $30$\,days after the initial alert. As a result, our catalog includes $60$ alerts (from GCN) likely associated with GRBs that occurred in the month leading up to May 14. In contrast, the follow-up window for TDEs is set to one year, leading to the inclusion of 6 TDEs (from TNS) in the catalog. The list was monitored throughout the observation period to identify all sources of interest that were potentially observable, based on geometric visibility constraints. Additionally, it was used to track which sources actually crossed the CT’s FoV during that time. We present a breakdown of the source list in \autoref{tab:full-source-table}.

\begin{table}[th]
    \centering
    \caption{Summary of the source list used for ToO follow-up searches. The total number of sources in the catalog, accounting for a maximum alert age of one year (except for steady sources), is provided in the "All sources" column. We also list all the sources that could have been observed ("Observable") and the sources that crossed the FoV ("FoV") during night 1 and night 2.}
    \label{tab:full-source-table}
    \resizebox{\textwidth}{!}{
    \begin{tabular}{|c|c|c|c|c|c|}\hline
         \textbf{Alert System} & \textbf{All Sources} & \textbf{Observable (night 1)} & \textbf{FoV (night 1)} & \textbf{Observable (night 2)} & \textbf{FoV (night 2)} \\\hline\hline
         GCN    & 1224 & 28 & 0 & 34 & 5 \\
         TNS    & 670  & 53 & 0 & 43 & 7 \\
         Steady & 120  & 80 & 0 & 68 & 11 \\
         Other  & 16   & 12 & 0 & 12 & 1 \\\hline
         \textbf{Total} & \textbf{2030} & \textbf{173} & \textbf{0} & \textbf{157} & \textbf{24} \\\hline
    \end{tabular}
    }
\end{table}

\begin{figure}[tpb]
    \centering
    \includegraphics[width=\textwidth]{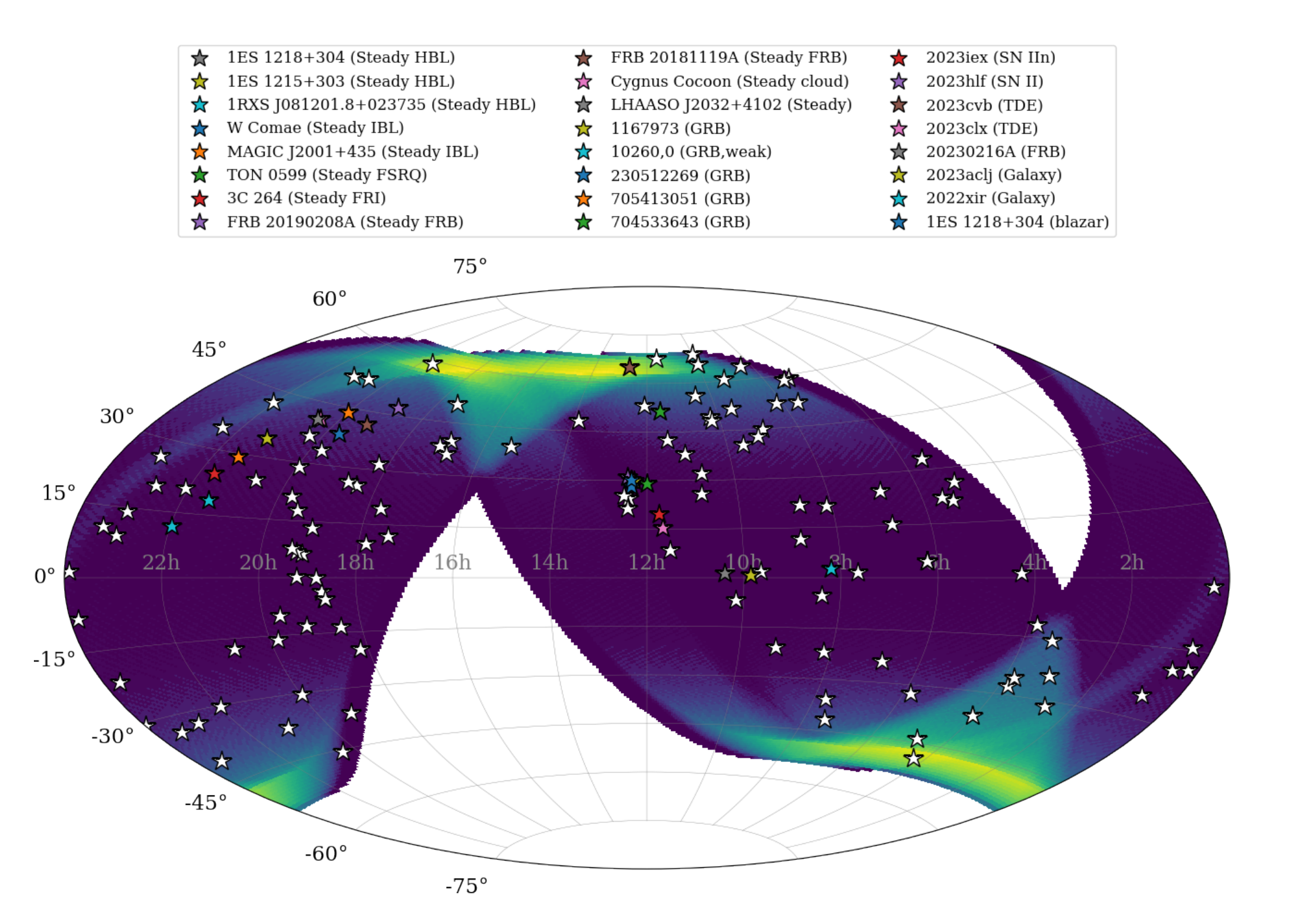}    
    \caption{Skymap of observable sources during night 2. White stars indicate all sources that could have been observed. Colored stars are sources that crossed the CT's FoV, and are identified in the legend. The background color scale is a measure of the maximum observation time for observable regions of the sky (more detail in the text).}
    \label{fig:observable-sources-night-1}
\end{figure}

As a first step, we track the sources' movement with respect to the detector by using NuTS to calculate their elevation and azimuth angles in the detector's reference frame at 1\,s intervals. We fold this information with pointing direction of the CT, which we determine by linearly interpolating between status messages that were reported every $\sim30$\,s over the course of the flight. Similarly, we use the tilt information provided during the flight. This allows us to relate the pointing direction with the locations of the sources in the sky and to calculate which sources were in the FoV at a given time. This procedure does not account for a threshold on Moon illumination, as data were collected during multiple periods without applying such a constraint. Following this procedure, we found a total of 24 sources (0 during night 1 and 24 during night 2) that entered and crossed the FoV while the CT was operational. Unfortunately, no observable source candidates for ToO follow-up searches were identified on night 1. During night 2, $11$ steady sources were identified, and more importantly $13$ transient sources (thus with higher priority) crossed the FoV. In \autoref{fig:observable-sources-night-1}, we show sky maps of the sources for night 2. The background color scale indicates the maximum observation time across all sky directions, serving as a proxy for the acceptance. This maximum observation time incorporates both FoV and Sun constraints; however, as previously noted, Moon illumination thresholds were not taken into account. Values are normalized to the maximum and displayed on a logarithmic scale ranging from $-1$ to $0$.

Amoung the sources that crossed the FoV during night 2, we find TDEs, GRBs, a flaring blazar, SNe, a FRB, and transient emissions from galaxies. Below, we provide a description of a selection of the GRB alerts. 1167973 corresponds to an alert of the Neil Gehrels Swift Observatory \cite{swift-alert} on May 10, 2023, and 705413051 corresponds to a Fermi Gamma-ray Burst Monitor (GBM) alert \cite{Fermi-alert} of the same source, classified as a long GRB, GRB 230510A. 230512269 corresponds to a Fermi-LAT alert associated with the bright short GRB 230512A. 704533643 corresponds to an older Fermi-GBM alert of the short GRB 230430A. The alert with the trigger number 10260,0 corresponds to a weak detection by Integral on May 10, 2023, with the source classified as a potential GRB. GRBs are compelling source candidates for the production of VHE neutrinos due to their large energy reservoirs and their non-thermal emissions, which demonstrate their ability to accelerate leptons to high energies, and thus makes them good potential sources of cosmic rays. Consequently, a variety of studies model their emissions of ultra-high energy cosmic rays and VHE neutrinos, for instance \cite{Waxman:1997ti, Murase:2007yt} (see also \cite{Guepin:2022qpl} and references herein). In addition, of all the sources that crossed the CT's FoV during EUSO-SPB2's flight, GRB 230510A had one of the highest priority ranking in our selected prioritization scheme (see \autoref{sec:NuTS}). GRB 230510A is therefore an interesting candidate for VHE neutrino searches, and in the following we focus on analyzing our observations of this source.

\subsection{Simulation criteria}

In order to calculate upper limits of the neutrino fluence of a transient source, and specifically of GRB 230510A, we follow the procedure presented below, which can be used for any source that crossed the CT's FoV during the flight. Due to the rapid rotation of the payload during descent on night 2, GRB 230510A spent only a total of 35\,s within the FoV. 

To begin, the source's elevation and azimuth angles as functions of time are passed to NuSpaceSim \cite{2023ICRCKrizmanic} to generate Monte Carlo $\nu_{\tau}$ event trajectories and sample $\tau$-lepton emergence probabilities and energy distributions that we will use as part of a simulation chain to calculate the CT'S acceptance to $\nu_{\tau}$s. While NuSpaceSim can also simulate EASs and propagate their Cherenkov signals to be scored by a basic detector model, we use separate tools for these steps to allow for higher-fidelity detector performance modeling (see also the discussion in ref. \cite{ICRC2025HeibgesDiffuse}). We use the event parameters from NuSpaceSim (e.g., event trajectory, sampled $\tau$-lepton energy and decay length) to sample an EASCherSim \cite{EASCherSim} shower library of upward-going EASs and generate Cherenkov light distributions. We then use a simplified detector model that accounts for wavelength-dependent response of the CT to make a first cut of the simulated events. We then further process passing events with the OffLine detector simulation \cite{JEM-EUSO:2023fyg}, which is our most robust representation of the detector response. 

The results of the detector simulations are tested against the event cuts developed for the EUSO-SPB2 CT's data analysis \cite{HeibgesPhDThesis2025}. These include cuts based on expectations of the bifocal shape and timing of the signals from simulations. We find that the acceptances calculated using the full NuSpaceSim simulation and our simulation chain using the simplified detector model described above agree well, whereas the simulation chain using the OffLine predicts about a factor of 10 lower acceptance. This is likely due to the enhanced treatment of detector effects (e.g., splitting the Cherenkov signal between pixels, accounting for losses in gaps between pixels) that the more detailed OffLine detector model provides.

\subsection{Upper limits and comparison to models}
\begin{figure}[t]
    \centering
    \includegraphics[width=0.95\linewidth]{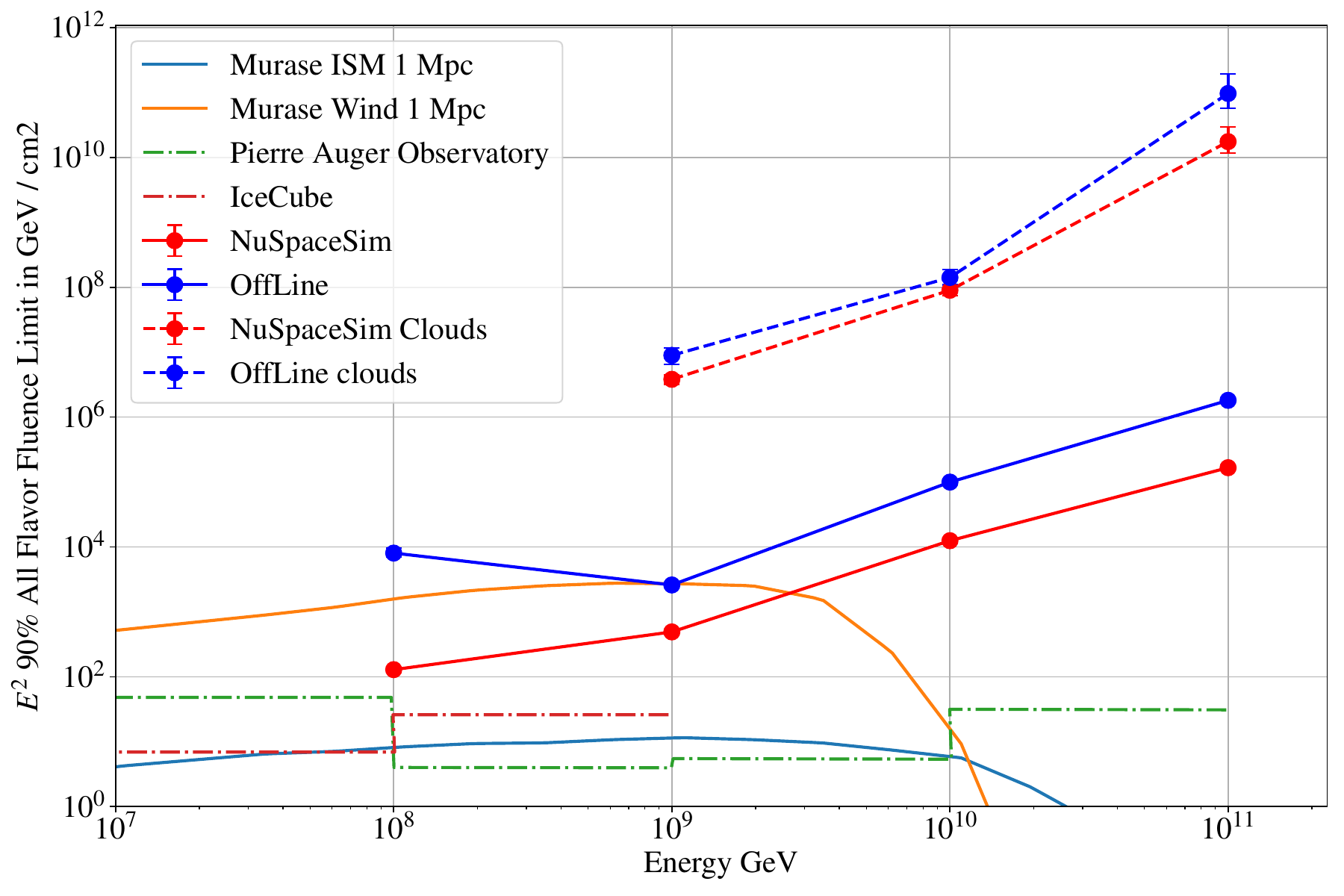}
    \caption{Calculated EUSO-SPB2 best-case upper limits on the neutrino fluence from GRB 230510A, assuming that the VHE neutrino burst occurred while the source was crossing the CT’s FoV. We show upper limits assuming no clouds (red) and accounting for clouds at $12$\,km (blue). For comparison, we show GRB fluence models from \cite{Murase:2007yt}, placing the source at a distance of $1$\,Mpc (blue and orange lines). For reference, the 90\% confidence upper limits from IceCube (green) and PAO (orange) in a $\pm$500\,s window around the gravitational wave event from GW170817 adapted from \cite{Ackermann:2022rqc} are shown.}
    \label{fig:too-sensitivity-vs-model}
\end{figure}

An analysis of the data taken from below-the-limb observations performed during the EUSO-SPB2 flight yielded no VHE neutrino candidates (see also ref. \cite{ICRC2025HeibgesDiffuse}). Thus, we calculate all-flavor upper limits on the neutrino fluence from our selected source candidate under the assumption of zero signal and zero background, as described in refs. \cite{Venters:2019xwi, HeibgesPhDThesis2025}. For this calculation, we assume a hypothetical scenario in which the VHE neutrino burst occurred while the source was crossing the CT’s FoV, despite GRB 230510A having occurred four days prior to the observation period. The upper limits presented therefore represent the detector’s optimal performance best-case sensitivity under these ideal conditions. The results are shown in \autoref{fig:too-sensitivity-vs-model}.

The calculated upper limits are approximately a factor of $\sim 100$ lower than our initial estimates \cite{ICRC2023Heibges}, which is explained by a different assumption for the observation time and for the PE detection threshold. The observation time initially assumed was 1000\,s, which is about $28$ times more than the $35$\,s of the source seen in EUSO-SPB2 during night 2. The PE threshold initially assumed was $\sim 20$ instead of the 47\,PEs assumed here. An important factor in this calculation is the presence of clouds during the night 2 observations. Based on cross-referencing with the MERRA2 cloud database, the cloud coverage is estimated to be homogenous and uniform at an altitude of $\sim12\,$km (for more details, see \cite{ICRC2025HeibgesDiffuse}). To account for the cloud cover in the simulations, a conservative estimate was made that the Earth-emerging tau-lepton has to decay above the cloud level. This high cloud cover, combined with the detector losing altitude to around 16\,km at the time of this observation, dropped the sensitivity by about 4 orders of magnitude.

We compare our upper limits for GRB 230510A with neutrino fluence predictions, using the models outlined in \cite{Murase:2007yt}. We chose a long GRB model, as GRB 230510A was classified as a long GRB. Two variations of this model are considered. The first variation (Murase ISM in \autoref{fig:too-sensitivity-vs-model}) uses the interstellar medium as a target for VHE neutrino production, and the second variation (Murase Wind in \autoref{fig:too-sensitivity-vs-model}) uses a denser photon target produced by the ejecta of the GRB progenitor. With these models, we determine the VHE neutrino horizon distance, which is the distance at which the modeled source fluence would result in one neutrino that would be detectable by EUSO-SPB2. We obtain $d_{\text{horizon}} \approx 0.28\,\text{Mpc}$ and $d_{\text{horizon}} \approx 4.29\,\text{Mpc}$ respectively (neglecting clouds). For a more realistic GRB distance at redshift $z=1$, the expected number of neutrinos seen in this observation of GRB 230510A are $N_{\nu,1,\text{clouds}} =  2.07^{+0.76}_{-0.49} \cdot 10^{-13}$, $N_{\nu,2,\text{clouds}} = 4.72^{+1.74}_{- 1.12} \cdot 10^{-11}$, $N_{\nu,1,\text{no clouds}} = 9.06^{+0.65}_{-0.55} \cdot 10^{-10}$ and $N_{\nu,2,\text{noclouds}} = 2.08^{+ 0.15}_{-0.13} \cdot 10^{-7}$.

\section{Conclusion}
The EUSO-SPB2 mission placed a Cherenkov detector in a near-space configuration on a super pressure balloon. We showed that the instrument was functional and able to observe EAS signatures. Furthermore, we demonstrated our ability to tilt and rotate the CT, allowing ToO follow-up observations of candidate neutrino sources. During our flight, we generated a list of possible neutrino sources, tilted the telescope below the limb, and performed a series of observations. The analysis of possible ToO observations during the descent of the balloon is ongoing. While the abbreviated flight of EUSO-SPB2 cut our observation time and data collection very short, the framework developed for this project can be used independently and applied to  future missions.\vspace{0.5cm}

\small{\noindent{\bf Acknowledgements} -- The authors acknowledge the support by NASA awards 11-APRA-0058, 16-APROBES16-0023, 17-APRA17-0066, NNX17AJ82G, NNX13AH54G, 80NSSC18K0246, 80NSSC18K0473, 80NSSC19K0626,  80NSSC18K0464,  80NSSC22K1488,  80NSSC19K0627 and 80NSSC22K0426, the French space agency CNES, National Science Centre in Poland grant n. 2017/27/B/ST9/02162, and by ASI-INFN agreement n. 2021-8-HH.0 and its amendments. This research used resources of the US National Energy Research Scientific Computing Center (NERSC), the DOE Science User Facility operated under Contract No. DE-AC02-05CH11231. We acknowledge the NASA BPO and CSBF staffs for their extensive support. We also acknowledge the invaluable contributions of the administrative and technical staffs at our home institutions.}

{\small
\bibliographystyle{JHEP-nt}
\bibliography{bibliography}}

\providecommand{\href}[2]{#2}\begingroup\raggedright\begin{thebibliography}{10}

\bibitem{IceCube:2018dnn}
M.G.~Aartsen et~al. ~\href{https://doi.org/10.1126/science.aat1378}{\emph{Science} {\bfseries 361} (2018) eaat1378} [\href{https://arxiv.org/abs/1807.08816}{{\ttfamily 1807.08816}}].

\bibitem{IceCube:2018cha}
{\scshape IceCube} collaboration ~\href{https://doi.org/10.1126/science.aat2890}{\emph{Science} {\bfseries 361} (2018) 147} [\href{https://arxiv.org/abs/1807.08794}{{\ttfamily 1807.08794}}].

\bibitem{2019ATel12967....1T}
I.~{Taboada} and R.~{Stein} ~{\emph{The Astronomer's Telegram} {\bfseries 12967} (2019) }.

\bibitem{Giommi:2020viy}
P.~Giommi, P.~Padovani, F.~Oikonomou et~al. ~\href{https://doi.org/10.1051/0004-6361/202038423}{\emph{Astron. Astrophys.} {\bfseries 640} (2020) L4} [\href{https://arxiv.org/abs/2003.06405}{{\ttfamily 2003.06405}}].

\bibitem{Stein:2020xhk}
R.~Stein et~al. ~\href{https://doi.org/10.1038/s41550-020-01295-8}{\emph{Nature Astron.} {\bfseries 5} (2021) 510} [\href{https://arxiv.org/abs/2005.05340}{{\ttfamily 2005.05340}}].

\bibitem{Reusch:2021ztx}
S.~Reusch et~al. ~\href{https://doi.org/10.1103/PhysRevLett.128.221101}{\emph{Phys. Rev. Lett.} {\bfseries 128} (2022) 221101} [\href{https://arxiv.org/abs/2111.09390}{{\ttfamily 2111.09390}}].

\bibitem{vanVelzen:2021zsm}
S.~van Velzen et~al. ~\href{https://doi.org/10.1093/mnras/stae610}{\emph{Mon. Not. Roy. Astron. Soc.} {\bfseries 529} (2024) 2559} [\href{https://arxiv.org/abs/2111.09391}{{\ttfamily 2111.09391}}].

\bibitem{IceCube:2022der}
{\scshape IceCube} collaboration ~\href{https://doi.org/10.1126/science.abg3395}{\emph{Science} {\bfseries 378} (2022) 538} [\href{https://arxiv.org/abs/2211.09972}{{\ttfamily 2211.09972}}].

\bibitem{IceCube:2023ame}
{\scshape IceCube} collaboration ~\href{https://doi.org/10.1126/science.adc9818}{\emph{Science} {\bfseries 380} (2023) adc9818} [\href{https://arxiv.org/abs/2307.04427}{{\ttfamily 2307.04427}}].

\bibitem{KM3NeT:2025npi}
{\scshape KM3NeT} collaboration ~\href{https://doi.org/10.1038/s41586-024-08543-1}{\emph{Nature} {\bfseries 638} (2025) 376}.

\bibitem{Venters:2019xwi}
T.M.~Venters, M.H.~Reno, J.F.~Krizmanic et~al. ~\href{https://doi.org/10.1103/PhysRevD.102.123013}{\emph{Phys. Rev. D} {\bfseries 102} (2020) 123013} [\href{https://arxiv.org/abs/1906.07209}{{\ttfamily 1906.07209}}].

\bibitem{ICRC2023Heibges}
{\scshape JEM-EUSO} collaboration ~\href{https://doi.org/10.22323/1.444.1134}{\emph{PoS} {\bfseries ICRC2023} 1134} [\href{https://arxiv.org/abs/2310.12310}{{\ttfamily 2310.12310}}].

\bibitem{Bagheri:2024byu}
M.~Bagheri et~al. ~\href{https://doi.org/10.1016/j.nima.2024.169999}{\emph{Nucl. Instrum. Meth. A} {\bfseries 1070} (2025) 169999} [\href{https://arxiv.org/abs/2406.08274}{{\ttfamily 2406.08274}}].

\bibitem{Adams:2025owi}
J.H.~Adams et~al. ~{\emph{arXiv e-prints} (2025) } [\href{https://arxiv.org/abs/2505.20762}{{\ttfamily 2505.20762}}].

\bibitem{ICRC2025Eser}
J.~{Eser}, A.V.~{Olinto} and G.~{Osteria} ~\href{https://doi.org/10.48550/arXiv.2509.04302}{\emph{arXiv e-prints} (2025) } [\href{https://arxiv.org/abs/2509.04302}{{\ttfamily 2509.04302}}].

\bibitem{ICRC2025HeibgesDiffuse}
T.~Heibges, D.~Garg, C.~Gu\'epin et~al. ~{\emph{PoS} {\bfseries ICRC2025} 1155}.

\bibitem{Guepin:2022qpl}
C.~Gu\'epin, K.~Kotera and F.~Oikonomou ~\href{https://doi.org/10.1038/s42254-022-00504-9}{\emph{Nature Rev. Phys.} {\bfseries 4} (2022) 697} [\href{https://arxiv.org/abs/2207.12205}{{\ttfamily 2207.12205}}].

\bibitem{2023ICRCWistrand}
H.~{Wistrand} et~al. ~\href{https://doi.org/10.22323/1.444.1185}{\emph{PoS} {\bfseries ICRC2023} 1185}.

\bibitem{blaufuss2019generation}
{\scshape IceCube} collaboration ~\href{https://doi.org/10.22323/1.358.1021}{\emph{PoS} {\bfseries ICRC2019} 1021} [\href{https://arxiv.org/abs/1908.04884}{{\ttfamily 1908.04884}}].

\bibitem{Heibges:2025llt}
T.~Heibges, C.~Gu{\'e}pin, L.~Kupari et~al. ~{\emph{arXiv e-prints} (2025) } [\href{https://arxiv.org/abs/2509.13844}{{\ttfamily 2509.13844}}].

\bibitem{swift-alert}
R.A.J.~{Eyles-Ferris}, P.A.~{Evans}, J.D.~{Gropp} et~al. ~{\emph{GRB Coordinates Network} {\bfseries 33752} (2023) }.

\bibitem{Fermi-alert}
C.~{Malacaria}, C.~{Meegan} and {Fermi GBM Team} ~{\emph{GRB Coordinates Network} {\bfseries 33766} (2023) }.

\bibitem{Waxman:1997ti}
E.~Waxman and J.N.~Bahcall ~\href{https://doi.org/10.1103/PhysRevLett.78.2292}{\emph{Phys. Rev. Lett.} {\bfseries 78} (1997) 2292} [\href{https://arxiv.org/abs/astro-ph/9701231}{{\ttfamily astro-ph/9701231}}].

\bibitem{Murase:2007yt}
K.~Murase ~\href{https://doi.org/10.1103/PhysRevD.76.123001}{\emph{Phys. Rev. D} {\bfseries 76} (2007) 123001} [\href{https://arxiv.org/abs/0707.1140}{{\ttfamily 0707.1140}}].

\bibitem{2023ICRCKrizmanic}
J.F.~{Krizmanic} et~al. ~\href{https://doi.org/10.22323/1.444.1110}{\emph{PoS} {\bfseries ICRC2023} 1110}.

\bibitem{EASCherSim}
A.~Cummings, R.~Aloisio, J.~Eser and J.~Krizmanic ~\href{https://doi.org/10.1103/PhysRevD.104.063029}{\emph{Phys. Rev. D} {\bfseries 104} (2021) 063029} [\href{https://arxiv.org/abs/2105.03255}{{\ttfamily 2105.03255}}].

\bibitem{JEM-EUSO:2023fyg}
{\scshape JEM-EUSO} collaboration ~\href{https://doi.org/10.1088/1748-0221/19/01/P01007}{\emph{JINST} {\bfseries 19} (2024) P01007} [\href{https://arxiv.org/abs/2309.02577}{{\ttfamily 2309.02577}}].

\bibitem{HeibgesPhDThesis2025}
T.~Heibges, \emph{{First Results from the Cherenkov Telescope Flown on EUSO-SPB2}}, Ph.D. thesis, {{Colorado School of Mines}}, 2025.

\bibitem{Ackermann:2022rqc}
M.~Ackermann et~al. ~\href{https://doi.org/10.1016/j.jheap.2022.08.001}{\emph{JHEAp} {\bfseries 36} (2022) 55} [\href{https://arxiv.org/abs/2203.08096}{{\ttfamily 2203.08096}}].

\end{thebibliography}\endgroup

\clearpage

    \newpage
{\Large\bf Full Authors list: The JEM-EUSO Collaboration}

\begin{sloppypar}
{\small \noindent
M.~Abdullahi$^{ep,er}$              
M.~Abrate$^{ek,el}$,                
J.H.~Adams Jr.$^{ld}$,              
D.~Allard$^{cb}$,                   
P.~Alldredge$^{ld}$,                
R.~Aloisio$^{ep,er}$,               
R.~Ammendola$^{ei}$,                
A.~Anastasio$^{ef}$,                
L.~Anchordoqui$^{le}$,              
V.~Andreoli$^{ek,el}$
% ...
}
\end{sloppypar}
\vspace*{.3cm}

{ \footnotesize
\noindent
%
% ...
$^{cb}$ Universit\'e de Paris, AstroParticule et Cosmologie (CNRS), Paris, France\\
% ...
$^{ef}$ Istituto Nazionale di Fisica Nucleare (INFN), Sezione di Napoli, Naples, Italy\\
% ...
$^{ei}$ Istituto Nazionale di Fisica Nucleare (INFN), Sezione di Roma Tor Vergata, Rome, Italy\\
% ...
$^{ek}$ Istituto Nazionale di Fisica Nucleare (INFN), Sezione di Torino, Turin, Italy\\
$^{el}$ Universit\`a di Torino, Dipartimento di Fisica, Turin, Italy\\
% ...
$^{ep}$ Gran Sasso Science Institute (GSSI), L’Aquila, Italy\\
$^{er}$ Istituto Nazionale di Fisica Nucleare (INFN), Laboratori Nazionali del Gran Sasso, Assergi, Italy\\
% ...
$^{ld}$ University of Alabama in Huntsville, Huntsville, AL, USA\\
$^{le}$ City University of New York (CUNY), Lehman College, Bronx, NY, USA\\
% ...
}

\end{document}